\documentclass[aps,prl,preprint,groupedaddress]{revtex4-1}

\usepackage{amsmath}
\usepackage{amssymb}
\usepackage{graphicx}
\usepackage{mathrsfs}
\usepackage{lineno}
\usepackage{bbm}

\begin{document}

\title{A detailed model of gene promoter dynamics reveals the entry into productive elongation to be a highly punctual process.}

\author{Jaroslav Albert\\
	jaroslav.albert@ronininstitute.org}
\affiliation{Ronin Institute}

\begin{abstract}
\small
Gene transcription is a multistep stochastic process that involves thousands of reactions. The first set of these reactions, which happen near a gene promoter, are considered
to be the most important in the context of stochastic noise and have been a focus of much research. The most common models of transcription are primarily concerned with the effect of activators/repressors on the overall transcription rate and approximate the basal transcription processes as a one step event. According to such effective models, the Fano factor of mRNA copy distributions is always greater than (super-Poissonian) or equal to 1 (Poissonian), and the only way to go below this limit (sub-Poissonian) is via a negative feedback. It is partly due to this limit that the first stage of transcription is held responsible for most of the stochastic noise in mRNA copy numbers.
However, by considering all major reactions that build and drive the basal transcription machinery, from the first protein that binds a promoter to the entrance of the transcription complex (TC) into productive elongation, it is shown that the first two stages of transcription, namely the pre-initiation complex (PIC) formation and the promoter proximal pausing (PPP), is a highly punctual process. In other words, the time between the first and the last step of this process is narrowly distributed, which gives rise to sub-Poissonian distributions for the number of TCs that have entered productive elongation. In fact, having simulated the PIC formation and the PPP via the Gillespie algorithm using 2000 distinct parameter sets and 4 different reaction network topologies, it is shown that only 4.4$\%$ give rise to a Fano factor that is $>1$ with the upper bound of 1.7, while for 31$\%$ of cases the Fano factor is below 0.5, with 0.19 as the lower bound. The two parameters that exert the most control on the Fano factor are the binding and dissociation rates between the activator and its enhancer. In the limit as these rates become low, our model approaches the conventional two-state promoter model, which can give rise Fano factors much larger than one. These results cast doubt on the notion that most of the stochastic noise observed in mRNA distributions always originates at the promoter.  
\normalsize
\end{abstract}

\maketitle

\section*{Introduction}
Gene transcription is a stochastic process involving a multitude of biochemical reactions by which the genetic code of a gene is copied into messenger RNA (mRNA). 
In eukaryotes, transcription occurs in four stages: 1) pre-initiation complex (PIC) formation \cite{Luse,Greber, Petrenko}; 2) promoter proximal pausing (PPP) \cite{Chen,Jonkers,Mayer,Li,Levine,Core,Core2}; 3) productive elongation \cite{Gonzalez,Bai,Muniz}; and 4) termination \cite{Rosonina,Proudfoot,Porrua,Porrua2,Kuehner,Richard,Luo,Kornblihtt}.
In stage 1) general transcription factors (GTF) bind the promoter in order to recruit RNA Polymerase II (Pol II) and other GTFs that enable the Pol II complex to begin transcription. In stage 2) the Pol II complex pauses a short distance downstream from the promoter site in order to give time to necessary processes, such as RNA capping and the addition of elongation-promoting factors, to do their job. In stage 3) the Pol II complex - known at this stage as elongation complex (EC) - enters the main body of a gene where it continues to transcribe mRNA until stage 4) in which transcription is terminated and the completed mRNA, along with the Pol II complex, are released. There are additional stages, such as splicing and the synthesis of ribonucleoprotein particles \cite{Stefl}; however, these tend to occur (though not always) during the elongation stage and can be incorporated into the elongation process itself \cite{Bentley}. 

Traditionally, stochastic models of transcription tend to focus on a highly simplified version of stage 1), while either ignoring the subsequent stages \cite{Walczak} or handling them by introducing a delay \cite{Ribeiro}. By doing so, such models are forced to operate under the assumption that whatever noise in mRNA copy numbers they are meant to explain/simulate, it all comes from this first stage. For super-Poissonian distributions, these models do quite well, owing to the multitude of promoter states that can be produced via activators/repressors. On the other hand, for sub-Poissonian distributions, this class of models works only in the presence of a negative feedback and only under certain conditions \cite{Dublanche,Tao,Lago,Zhang1}. However, models that take into account the reinitiation of Pol II at the promoter have revealed that mRNA  distributions can in fact be sub-Poissonian without any feedback \cite{Karmakar}, or at the very least reduce the variance \cite{Boettiger,Liu,Filatova,Braichenko,Choubey,Choubey2,Xu,Cao} as compared to the above-mentioned simpler models. This discovery challenges the conventional wisdom that sub-Poissonian mRNA distributions are necessarily indicative of a negative feedback. 

In this paper, we present a detailed stochastic model of the PIC formation and the PPP, one that involves 15 macromolecules - 13 general transcription factors (GTF), 1 activator, and the RNA polymerase II. The model is based on the current understanding of these two transcription stages. The goal of this paper is to investigate the range of Fano factors for the number distribution of ECs in a given time. we do this by generating 500 parameter sets and simulate for each set the stochastic evolution of the system, starting with an empty promoter. A crucial feature of this model is 
the relationship between the PPP stage and the reinitiation of Pol II, which can take several forms, two of which are considered herein: 1) Pol II is able to bind the promoter only when the promoter proximal region is free of a paused complex (PC); and 2) Pol II can bind but it cannot clear the promoter while a PC is present. The effects of activator dynamics on the resulting Fano factors are also simulated using a simple on-off model, where ``on" means the activator is bound to the promoter and ``off" means it is not. Since the exact order in which the GTFs are recruited to the promoter is still unsettled, all simulations are repeated using models with one and two additional reaction pathways. The simulations show a remarkable reluctance of the Fano factors to rise above 1 when the activator dynamics are relatively fast. In those cases where
the binding rate is fast and the dissociation rate is slow, the Fano factors concentrate around the value of 0.4 with a maximum of 0.6. On the other hand, when both the binding and dissociation factors are low, the Fano factors can reach values in excess of xxx. While the correlation between the activator dynamics and the Fano factors is consistent with previous studies, the migration of the Fano factors
to such extremely low values is a novel feature of the highly sequential nature of stages 1) and 2).

\section*{Gene transcription}
\subsection*{Pre-initiation complex formation}

\begin{figure}
	\centering
	\includegraphics[trim=0 0 0 1.0cm, height=0.5\textheight]{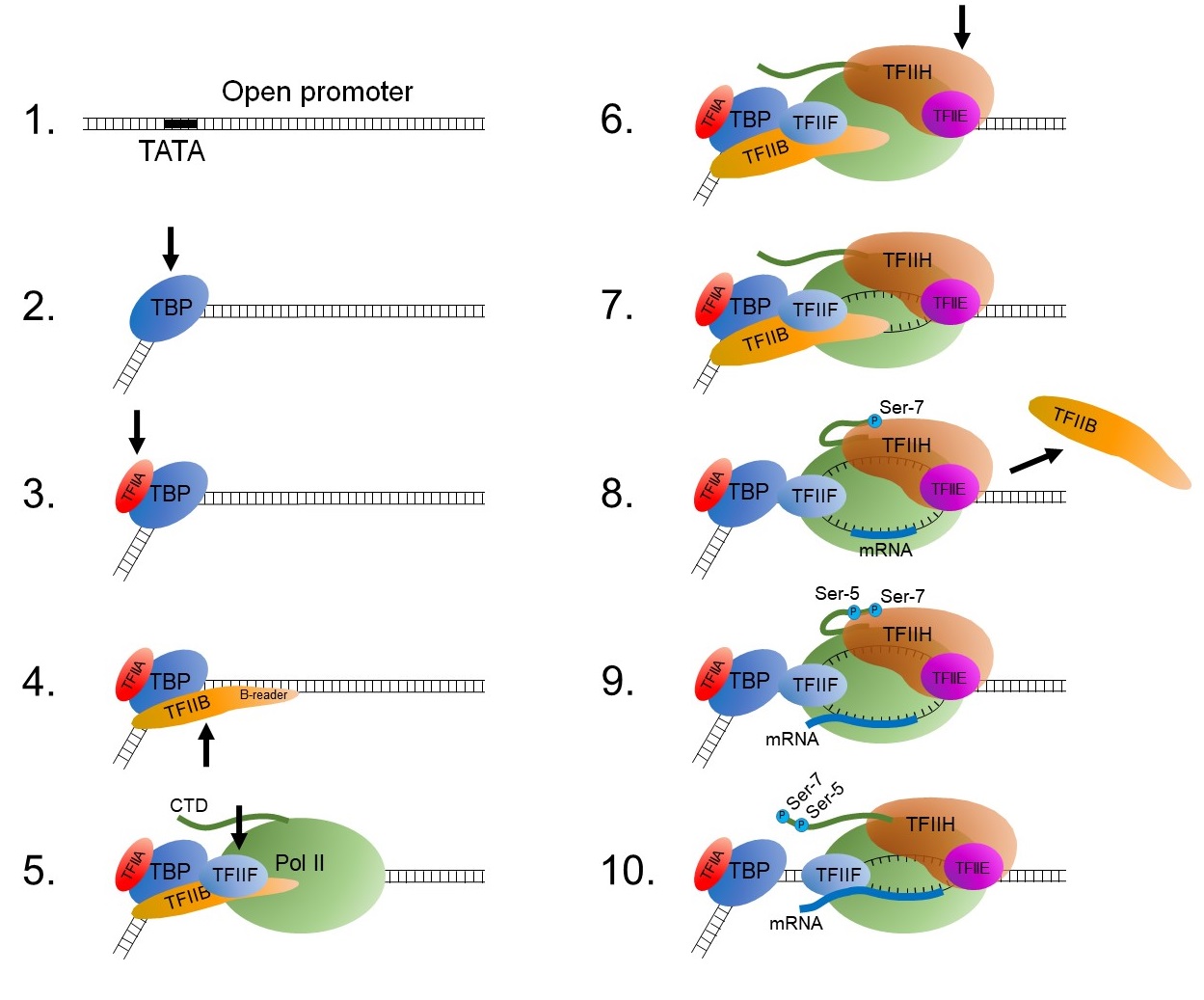}
	\caption{{\bf Pre-clearance complex (PCC) formation.} 1) Empty promoter; 2) TBP binds to TATA box and bends the DNA; 3) TFIIA binds TBP; 4) TFIIB binds the TATA/TBP complex;
	5) the complex Pol II/TFIIF is recruited; 6) the complex TFIIH/TFIIE is recruited; 7) initiation: DNA unwinds; 8) Serine 7 is phosphorelated, bubble grows and TFIIB is released; 
	9) Serine 5 is phosphorelated; 10) the EC clears the promoter.}\label{Fig1}
\end{figure}

The first step in mRNA transcription is the pre-initiation complex (PIC), the formation of which begins with the binding of a transcription factor (TF) called
TFIID to the promoter. TFIID is a multi sub-unit factor comprising of TBP (TATA binding protein) and TAFs (TBP associated factors). As the name suggests, the TBP binds to
the part of the promoter with the sequence TATA (Fig. \ref{Fig1}.1 and \ref{Fig1}.2). When the TBP binds the TATA box, it creates an almost 80 degree bend in the DNA. In order to stabilize this TATA-TBP complex, another TF, called TFIIA, binds
to it (Fig. \ref{Fig1}.3). Strictly speaking, TFIIA is not a general TF, as transcription can proceed without it, but merely functions as a scaffold protein.
The TF that is crucial for transcription is TFIIB (Fig. \ref{Fig1}.4), which can bind the promoter independently of TFIIA.
In this state, the promoter can receive the RNA polymerase II (Pol II). Since Pol II has no affinity to the DNA, it must bind the promoter via a TF called TFIIF (Fig. \ref{Fig1}.5), which has affinity to TFIIB. The TFIIB has a part that binds the promoter, and another part, called the B-reader, that ensured that Pol II faces the downstream direction.
Another role of TFIIF is to recruit TFIIE, which acts as a loading factor for another TF called TFIIH. TFIIE and TFIIH usually bind the promoter as a complex (Fig. \ref{Fig1}.6).
In higher eukaryotes -- which is the focus of this paper -- all these processes are happening in a 3-dimensional confirmation with the help of long range enhancers and mediator proteins, which together form the enhancer-mediator complex. For simplicity, the mediator or the enhancers were are not shown in Fig. \ref{Fig1}.

\subsection*{Initiation}
At this point the DNA is still closed and the Pol II is not yet making any mRNA. However, soon after TFIIH comes into play, the DNA unwinds in the activity center of Pol II (Fig. \ref{Fig1}.7). What causes this unwinding is a tug of war between 
TFIIH and TBP. Pol II is a processive enzyme and wants to move forward. However, because it is bound to TFIIB via TFIIF, it is unable to move. The XPB-subunit of TFIIH, which is the translocate part of TFIIH, tracks the DNA forward, but because it is bound to Pol II via TFIIE, it, too, cannot move. As a result of these opposing forces, the DNA unwinds. Such unwound DNA is called a bubble \cite{Pal}. At this stage the Pol II tries to synthesize mRNA but is blocked by the B-reader part of TFIIB, which partially extends to the active site of Pol II. The partial blocking of the B-reader ensures that Pol II begins transcription at the correct site, called +1. Since this blocking is only partial, Pol II sometimes manages to produce short transcripts $\sim$ 3-5 base pair long, which, unless the Pol II can clear the promoter, are released from the active site -- a process known as abortive transcription (not shown in the figure). This transition from closed to open bubble is called initiation.

\subsection*{Promoter clearance}
When the nascent RNA grows to a length of about 9 nucleotides, thus strengthening the RNA-DNA bond, it becomes harder to abort. The growing transcript forces TFIIB to dissociate, allowing the bubble to grow to a length of about 18 nucleotides (Fig. \ref{Fig1}.8). However, such a large bubble is thermodynamically unstable and collapses to a smaller bubble of length of about 10 nucleotides. This collapse creates a force that causes a forward movement of Pol II.
At the time of the bubble collapse, the C-terminal domain (CTD) of Pol II is phosphorylated (Fig. \ref{Fig1}.8 and \ref{Fig1}.9). This phosphorylation is facilitated by the CDK7 subunit of TFIIH and the enhancer-bound kineses, which add phosphate groups to Serine 7 and Serine 5 (in that order) of each heptad repeat of the CDT.
Serine 5 phosphorylation breaks the bond between Pol II and the mediator complex, allowing Pol II to move forward and increasing the transcript to a length of 20 base pairs on average (Fig. \ref{Fig1}.10).

Note: by conventional definition, PIC refers to the protein complex consisting of TBP, TFIIA (optional), TFIIB, TFIIE, TFIIF, Pol II and TFIIH. For convenience we will define the term ``Pre-clearance complex" (PCC) to mean the state of the Pol II complex prior to clearance in order to place the initiation and phosphorylation processes under one umbrella.  

\subsection*{Promoter proximal pausing and escape}

\begin{figure}
	\centering
	\includegraphics[trim=0 0 0 1.0cm, height=0.5\textheight]{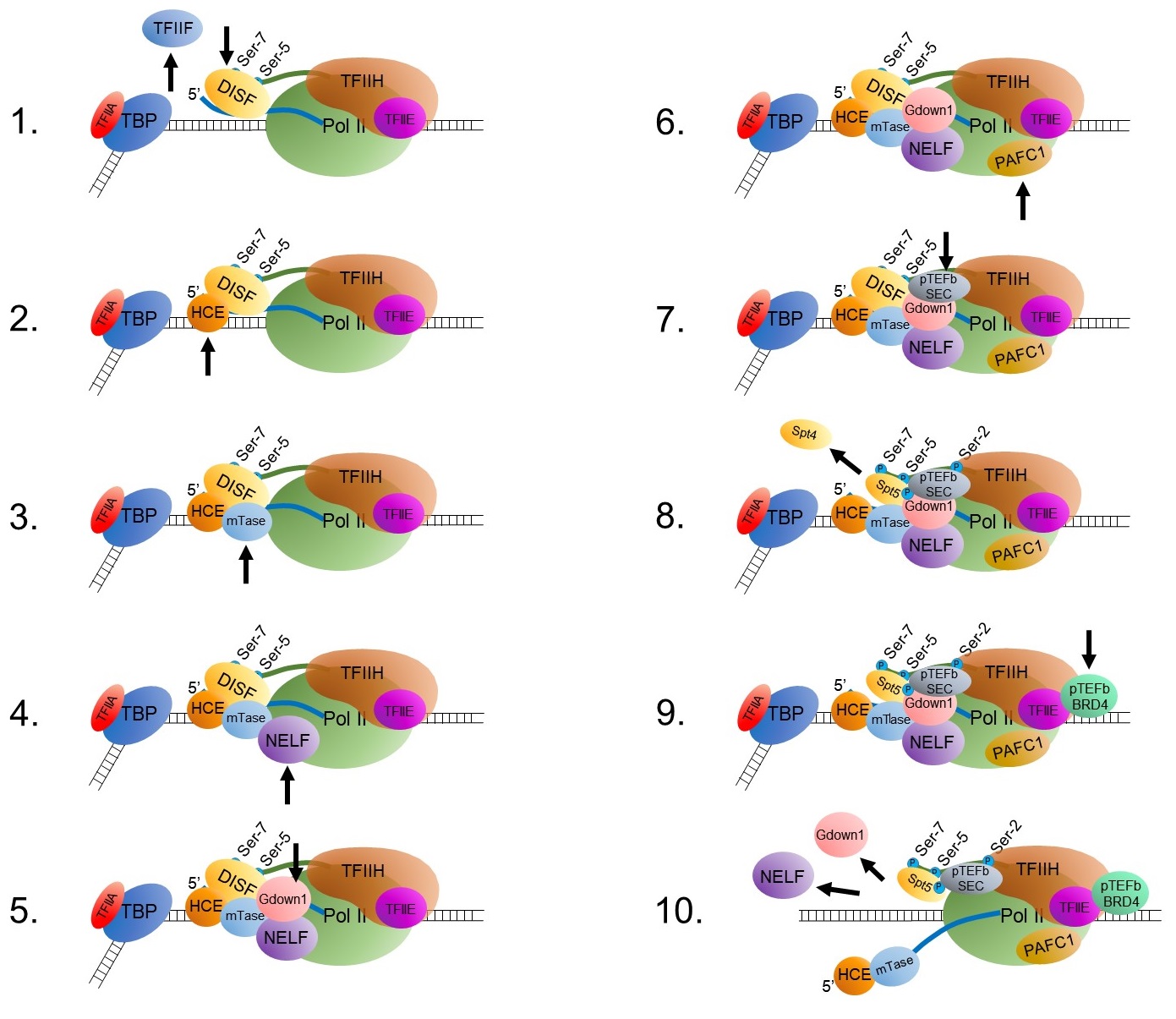}
	\caption{{\bf Promoter proximal pausing and escape.} 1) DISF is recruited by phosphorelated Serine-5 and Serine-7 to the nascent mRNA; 2) HCE caps the 5' end of mRNA; 3) mTASE is recruited to the 5' end; 4) NELF is recruited; 5) Gdown1 is recruited; 6) PAFC1 is recruited; 7) the pTEFb/SEC complex is recruited; 8) the pTEFb/SEC complex phosphorelates Serine 2; 9) the complex pTEFb/BRD is recrired; 10) escape into productive elongation.}\label{Fig2}
\end{figure}

The 5' end of the transcript is now sticking out of the Pol II exit channel. At this time, Spt 4 and Spt 5, which togeher are known as the DSIF complex, are recruited to the phosphorelated Serine-5 and Serine-7 of the CDK, while in the process removing TFIIF (Fig. \ref{Fig2}.1).
The function of the DSIF complex is to recruit two capping enzymes, the human capping enzyme (HCE) and mTASE, to the 5' end of the transcript (Fig. \ref{Fig2}.2 and \ref{Fig2}.3), which increases the stability of the RNA, and also to recruit a TF called NELF (negative elongation factor) (Fig. \ref{Fig2}.4). NELF
freezes Pol II, thus locking the RNA-DNA hybrid inside its active center. Another TF, called Gdown1, binds the growing complex (Fig. \ref{Fig2}.5) and helps NELF to maintain this frozen state. Gdown1 also prevents
TFIIF from binding back to Poll II and keeps protein called TTF2 (transcription termination factor 2) from binding the complex. Another TF called PAFC1 (Pol associated factor complex 1) binds Pol II (Fig. \ref{Fig2}.6); it has a dual role of promoting the frozen state and helps maintain elongation. 
To overcome this frozen state, two special proteins BRD-4 and SEC bind a TF called pTEF-b, thus stimulating its kinase activity. The pTEF-b/SEC complex changes the comformation of PAFC1 and phosphorelates the Spt5 part of DISF and Seren-2 of the CTD, which releases the Stp4 part of DISF in the process (Fig. \ref{Fig2}.8). Now the PAFC1 and Spt5, which were previously negative regulators in the promoter proximal state, are now positive regulators of elongation and stay bound to Pol II. To complete the transition from the PPP to productive elongation, pTEF-b/BRD-4 must bind the complex (Fig. \ref{Fig2}.9), which stimulates the release of Spt4, NELF and Gdown1, allowing Pol II to enter the elongation phase (Fig. \ref{Fig2}.10).

\subsection*{Reinitiation of Pol II}
When the Pol II complex clears the promoter, the promoter returns to one of the states TATA, TATA/TBP or TATA/TBP/A. That means that the PIC can begin to form again. However, due to structural reasons, e. g. the size of the PC (paused complex) and its location on the gene, a newly bound Pol II may not have enough physical space to clear the promoter \cite{Darst,Hahn,Schier,Adelman}. The question as to whether another Pol II can even bind the promoter in the presence of a PC or whether it is simply blocked from clearing has not been settled. However, there is increasing evidence for a negative correlation between the amount of time the PC stays in the paused state and the clearance rate \cite{Shao,Gressel}.


\section*{Materials and Methods}
\subsection*{The model}

\begin{figure}
	\centering
	\includegraphics[trim=0 0 0 1.0cm, height=0.5\textheight]{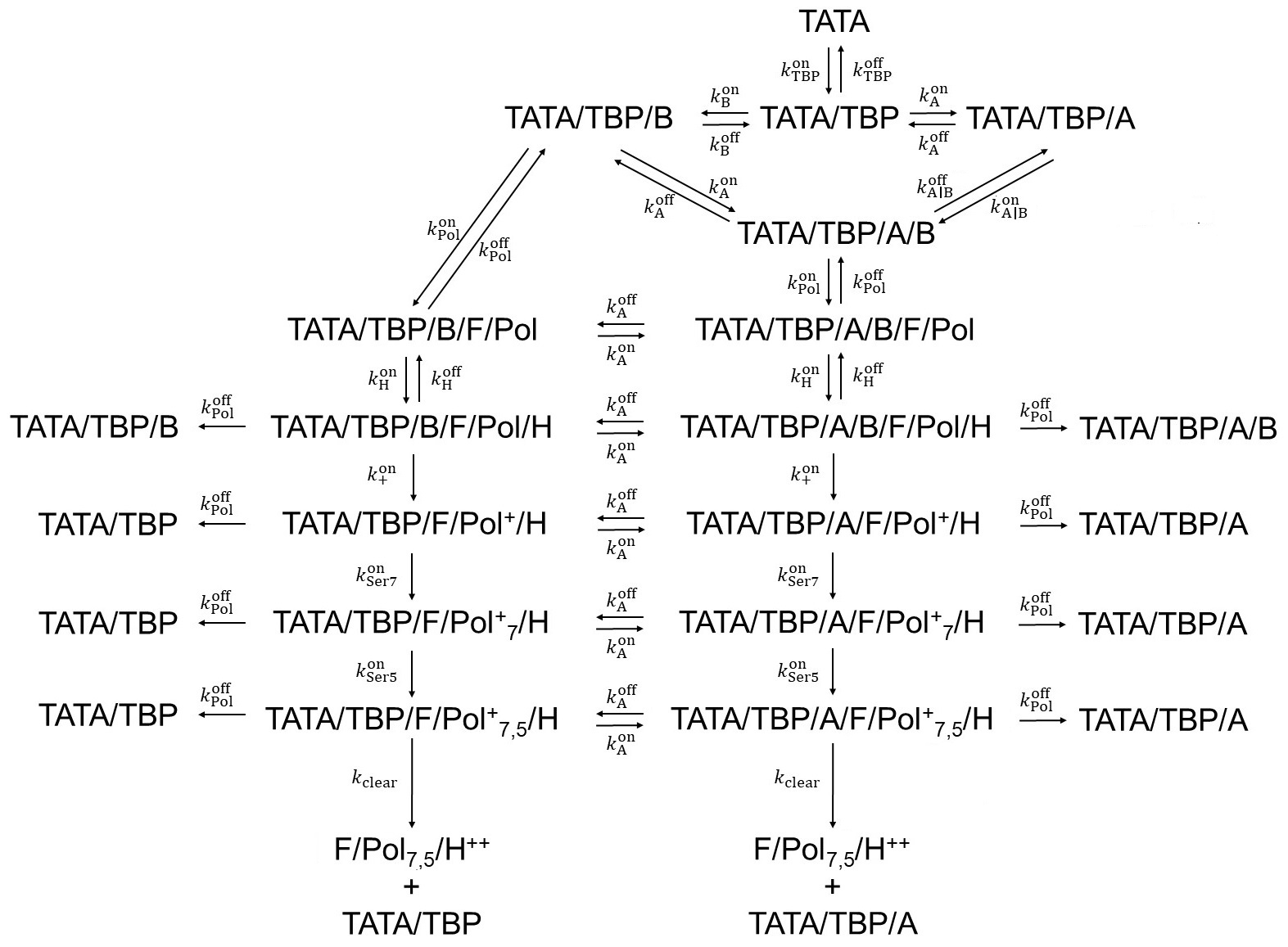}
	\caption{Reactions involved in the pre-clearance complex (PCC) formation.}\label{Fig3}
\end{figure}
\begin{figure}
	\centering
	\includegraphics[trim=0 0 0 1.0cm, height=0.5\textheight]{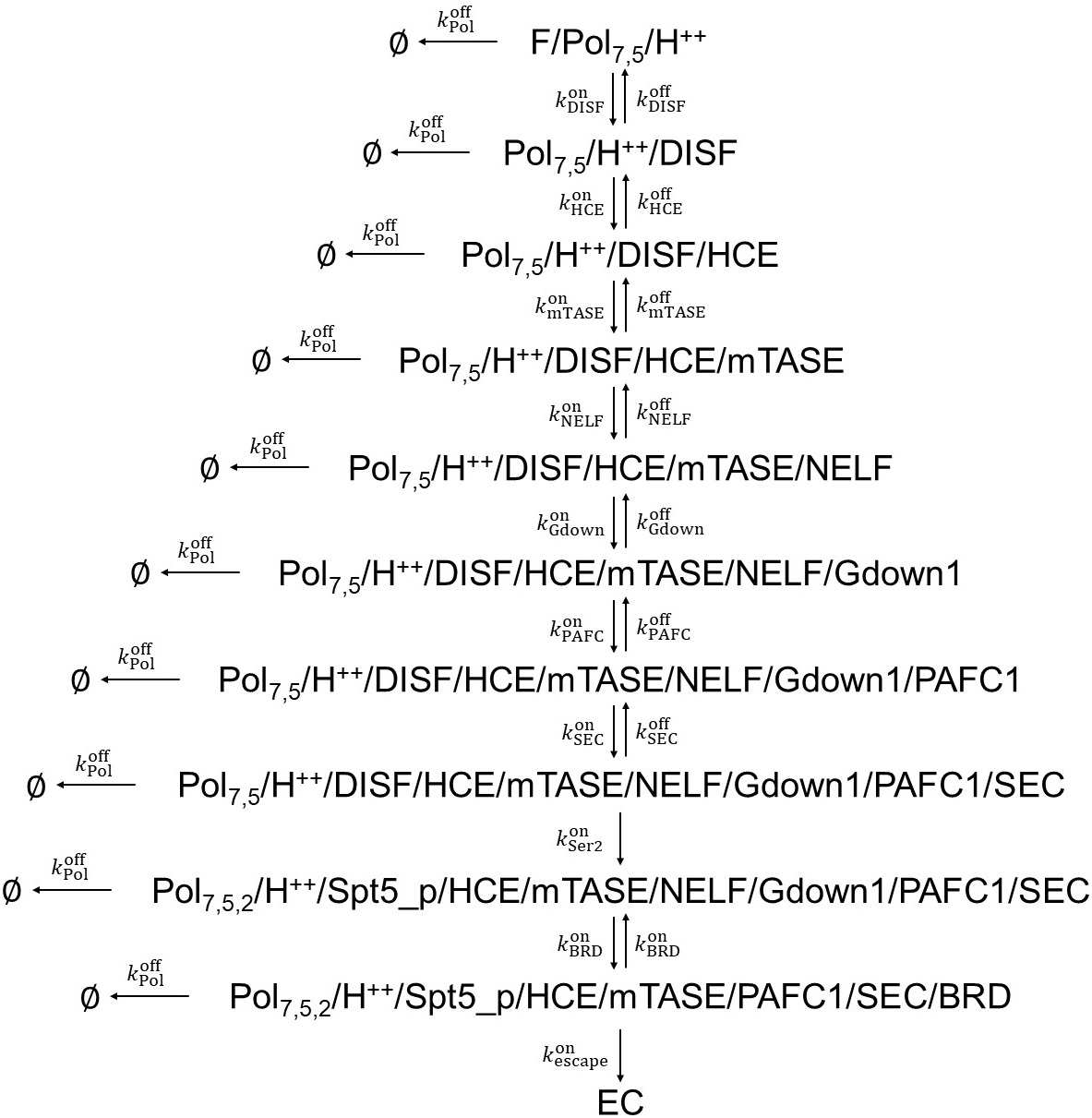}
	\caption{Reactions involved in the promoter proximal pausing and escape.}\label{Fig4}
\end{figure}
The biochemical reactions that form the PCC and those that cause and maintain the PC are shown in Fig. \ref{Fig3} and \ref{Fig4}, respectively. The variables, i. e. promoter states, are labeled so as to make clear what proteins are bound to the complex and what parts of the complex have been phosphorylated. 
For example, in the transition from TATA/TBP/B/F/Pol/H to TATA/TBP/F/Pol$^+$/H, the complex has moved into the initiation phase, conveyed by the plus sign, and TFII B was released in the process, as indicated by the absence of B in the latter state. This transition changes the value of the former variable from 1 to 0, and conversely so for the latter variable. The subscript in, e. g. TATA/TBP/F/Pol$^+_{7,5}$/H, indicates that both Serine 7 and Serine 5 have been phosphorylated. In Fig. 4, the double plus in the superscript of H signifies that the PCC has cleared the promoter.
The reactions that lead to the state $\emptyset$ cause Pol II to dissociate from the DNA. Hence, $\emptyset$, which also takes on the values of 0 or 1, represents
an empty DNA in the region downstream from the PCC. Two distinct interactions between the PCC and PC were considered: 1) Pol II cannot bind the promoter unless $\emptyset =1$; and 2) Pol II can bind the promoter but cannot clear it unless $\emptyset =1$.
Thus, the reaction propensity for case 1) was $k_{\text{Pol}}^{\text{on}}\emptyset$, while the propensity in case 2) was $k_{\text{clear}}\emptyset$.

In addition to the reactions in Figs. 3 and 4, the following reactions were added to the system:
\begin{eqnarray}\label{reactions}
	&&\emptyset\xrightarrow{\makebox[1cm]{$k^{\text{on}}_{\text{act}}$}}Z\nonumber\\
	&&Z\xrightarrow{\makebox[1cm]{$k^{\text{off}}_{\text{act}}$}}\emptyset,\nonumber\\
\end{eqnarray}
where $Z$ is the state of an enhancer, such that when $Z=1$, the enhancer is occupied by an activator, and when $Z=0$, the enhancer is unoccupied. Although the mechanism by which activators enhance
transcription is not unique and can occur in both the PCC formation \cite{Cosma} and during the PPP \cite{XavierChen, LJCore}, some evidence suggests that in the PCC formation the primary role of activators is to bring Pol II to the promoter \cite{Ma}.
To keep things simple, we have incorporated the state of the enhancer, i. e. the variable $Z$, into the main model by multiplying the propensities for transitioning from the state TATA/TBP/B and TATA/TBP/A/B to TATA/TBP/B/Pol and TATA/TBP/A/B/Pol, respectively, by $(1+99Z)/100$. The logic behind this factor is simple. When $Z=0$, and remains so (no activator, i. e. $k^{\text{on}}_{\text{act}}=0$), the propensity for the binding of Pol II is $k^{\text{on}}_{\text{Pol}}/100$, which diminishes the overall rate of transcription; we consider this as the state of basal transcription. When $Z=1$, and remains so, i. e. $k^{\text{off}}_{\text{act}}=0$, the propensity for the binding of Pol II is restored to $k^{\text{on}}_{\text{Pol}}$. Finally, there is the intermediate situation, in which 
$k^{\text{on}}_{\text{Pol}}\neq 0$ and $k^{\text{off}}_{\text{Pol}}\neq 0$. By regulating $k^{\text{on}}_{\text{Pol}}$ and $k^{\text{off}}_{\text{Pol}}$, one can not only adjust the overall transcription rate, but also the stochastic nose in the system, as will be demonstrated later.

\subsection*{Parameter selection}
The parameters used in all simulations were selected from ranges that are consistent with values found in literature. Table 1 shows the parameter ranges for the PIC formation and the relevant references. 
\begin{table}[htbp]
	\centering
	\caption{at $nM=8$}
\begin{tabular}{lllll}
	\hline
Description\,\,\,\,\,\,\,\,\,\,\,\,\,\,\,\,\,\,\,\,\,\,\,\,\,\,\,\,\,\,\,\,\,\,\,\,\,\,\,\,\,\,\,\,\,\,\,\,\,\,\,\,\,\,\,\,\,\,\,\,\,\,\,    & Symbol\,\,\,\,\,\,\,\,\,\,\,\,\,\,\,\,\,\,\,\,\,\, & Range\,\,\,\,\,\,\,\,\,\,\,\,\,\,\,\,\,\,\,\,\,\,\,\,\,\,\,\,\,\,\,\,\,\,\,\,\,\,\,\,\,\,\, & Reference \\
	\hline
		Binding rate of TBP             & $k_{\text{TBP}}$ & 0.001-0.1\,\text{s}$^{-1}$ & \cite{Chen,Zhang,Revyakin,Jung}  \\
		Binding rate of TFIIA            & $k_\text{A}$ & 0.01-0.1\,\text{s}$^{-1}$ & \cite{Zhang}  \\
		Binding rate of TFIIB            & $k_\text{B}$ & 0.025-0.25\,\text{s}$^{-1}$ & \cite{Zhang}  \\
		Binding rate of Pol II         & $k_{\text{Pol}}$ & 0.025-0.25\,\text{s}$^{-1}$ & \cite{Darzacq}   \\
	    Dissociation rate of TBP             & $k_{\text{TBP}}$ & 0.125-0.0125\,\text{s}$^{-1}$ & \cite{Zhang,Jung,Presman,Hasegawa,Heiss}  \\
	    Dissociation rate of TFIIA            & $k_{\text{A}}$ & 0.01-0.001\,\text{s}$^{-1}$ & \cite{Zhang,Jung}  \\
	    Dissociation rate of TFIIB            & $k_{\text{B}}$ & 0.6-0.006\,\text{s}$^{-1}$ & \cite{Zhang,Jung,Presman}  \\
	    Average pause time            & $t_\text{pause}$ & $\sim$5\,mins & \cite{Buckley,Boettiger}  \\
	    Efficiency            & $r$ & 5.5-83.3$\times 10^{-4}$\text{s}$^{-1}$ &   \\
	\hline
\end{tabular} 
\end{table}

For the PPP, specific reaction rates are much less available than the rates for the PIC formation. However, since the reactions that occur during the PPP are of the same kind as those in the PIC formation, namely association and dissociation of proteins, it is reasonable to assume they too have ranges similar to those given in Table 1, e. g. 0.001-0.1s$^{-1}$. Also, these rates can be constrained by the average duration of the pausing, for which there are data. In, for example, \cite{Buckley}, the average duration of the PPP, $\langle t_\text{pause}\rangle$, was measured to be $\sim$ 5 minutes. Note: in \cite{Buckley} and other studies, $\langle t_\text{pause}\rangle$ is not considered to be the time it takes the freshly cleared complex to reach the elongation stage; rather it is treated as the time that Pol II spends bound to the DNA during the paused state. In other words, $\langle t_\text{pause}\rangle$ is the average waiting time for either $\emptyset$ to go from 0 to 1, or 
for Pol$^{++}_{7,5,2}$/H/Spt5$_{\text{p}}$/HCE/mTASE/PAFC1/SEC/BRD to transition into PE. Another limiting factor that can help restrict not only the reaction rates of the PPP but the rates of the PCC formation as well, is the efficiency of transcription, or transcription rate $r$, which is typically in the range 5.5-83.3$\times 10^{-4}$\text{s}$^{-1}$. The rate $r$ is defined as the inverse of the average waiting time to reach productive elongation $\langle t_\text{elong}\rangle$, starting from randomized initial conditions, i. e. a value of one is randomly assigned to one of the variables TATA, TATA/TBP, TATA/TBP/A, TATA/TBP/B, TATA/TBP/A/B at time $t=0$. For the purposes of this paper it is not necessary to compute $\langle t_\text{elong}\rangle$ exactly; an estimate is sufficient and can be obtained by computing $\langle t_\text{clear}\rangle$, which is the average waiting time for clearance to occur, subject to the aforementioned randomized initial conditions, and then adding $\langle t_\text{escape}\rangle$, which is defined as the time to go from the state F/Pol$_{7,5}$/H to PE (see Fig. 4).  Thus, $r=\langle t_\text{elong}\rangle^{-1}$, where $\langle t_\text{elong}\rangle\sim \langle t_\text{clear}\rangle+\langle t_\text{escape}\rangle$. The exact details on how $\langle t_\text{clear}\rangle$, $\langle t_\text{escape}\rangle$ and $\langle t_\text{pause}\rangle$ were computed can be found in appendix A.

We generated 500 parameter sets in the range restriction $0.001-0.1$ and subject to the constraints $4<\langle t_\text{pause}\rangle<6$ min and $13\times 10^{-4}$s$^{-1}<r<20\times 10^{-4}$s$^{-1}$. Evidence suggests \cite{Zhang} that when TFIIA is bound to the promoter, the binding rate of TFIIB is significantly enhanced, while its dissociation rate is much reduced. This was taken into account by setting $k^{\text{on}}_{\text{A$|$B}}$ to $10k^{\text{on}}_{\text{B}}$ and $k^{\text{on}}_{\text{A$|$B}}$ to $0.1k^{\text{on}}_{\text{B}}$ (see Fig. 3). Note that while binding factors must have an upper bound, due to physical restrictions such as protein density, number of collisions with their target, et cetera, no such restrictions apply to the lower bounds of dissociation factors; they are instead restricted by thermodynamic stability and degradation by enzymatic reactions. Thus, the values of some dissociation factors below $10^{-3}$s$^{-1}$ were allowed. They were also necessary, as we will now explain.

The aforementioned scheme for computing $\langle t_\text{pause}\rangle$ and estimating $r$ was unable to generate $r$ in the desired range $13\times 10^{-4}$s$^{-1} - 20\times 10^{-4}$s$^{-1}$ when all parameter values were constrained to $0.001$s$^{-1}$ - $0.1$s$^{-1}$; the largest value of $r$ yielded by this range after ten thousand selections was $0.7\times 10^{-4}$s$^{-1}$. For this reason, we decided to lower the dissociation rate for Pol II in PCC (but not in the PC) by some factor $\beta$. Then, $\beta$ was incrementally increased until the scheme started producing the desired results. The lowest value of $\beta$ that yielded the desired results within ten thousand selections was 10. Thus, the range for the dissociation rate for Pol II in PCC was set to $0.0001 - 0.01$s$^{-1}$.

\subsection*{Simulations}
All simulations were done with the Gillespie algorithm (GA) \cite{Gillespie}. For all simulations, each realization in an ensemble started at time $t=0$, and with the initial conditions TATA$=1$, and stopped when the algorithm passed $t=60000$s. The number of times the reaction Pol$^{++}_{7,5,2}$/H/Spt5$_p$/HCE/mTASE/PAFC1/SEC/BRD$\rightarrow$ EC occurred before $t=60000$s was recorded. From now on, we will let EC represent this number. Note: the transcription rate $r$ is related to the average EC via $r=\langle EC\rangle/60000$. In all subsequent sections we will refer to $\langle EC\rangle$, rather than $r$.

\section*{Results}

\begin{figure}
	\centering
	\includegraphics[trim=0 0 0 1.0cm, height=0.8\textheight]{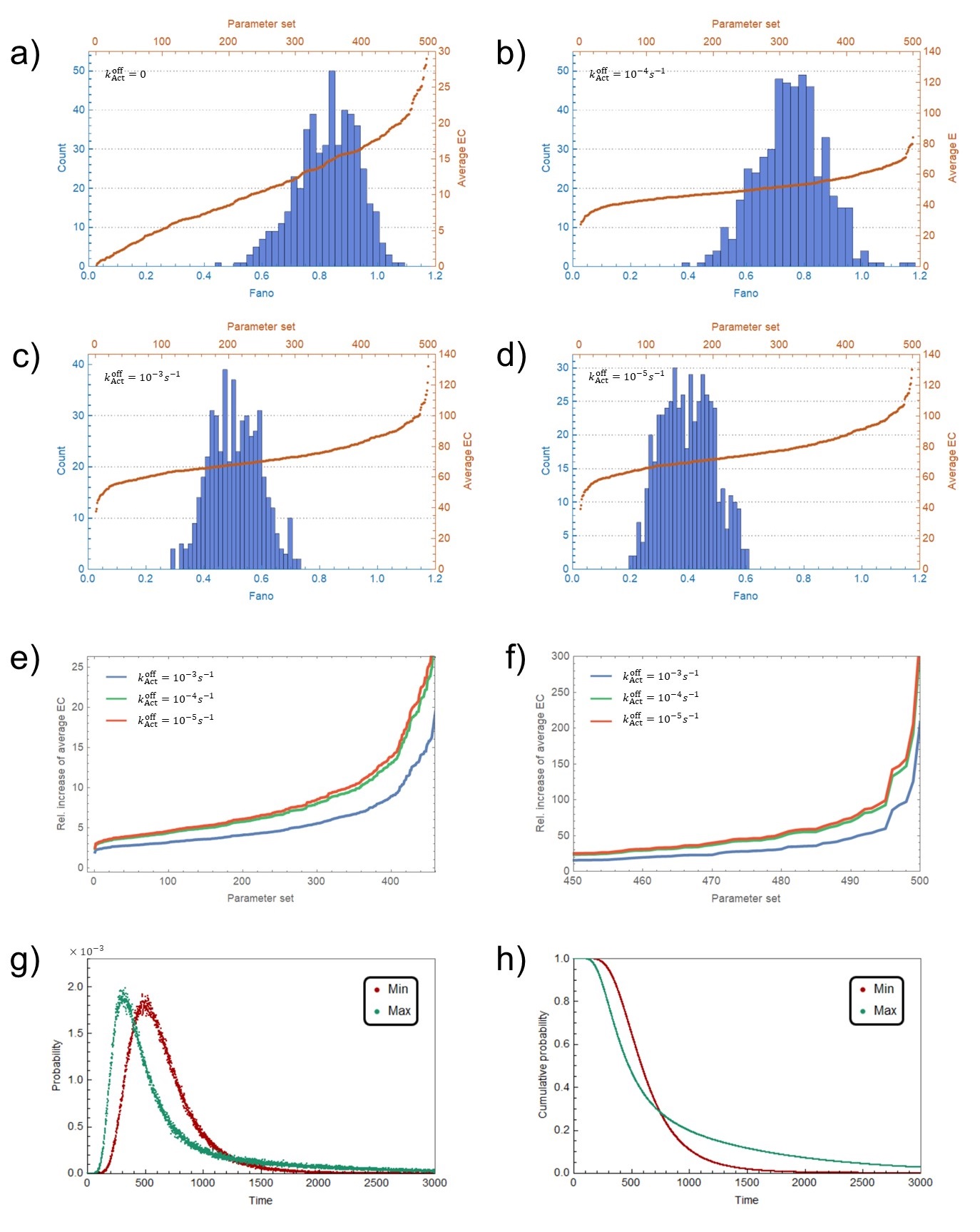}
	\caption{Results of simulations for PCC-PC interaction model 1. {\bf (a)} - {\bf (d):} Fano factors (blue bars) and average EC for all 500 parameter sets. {\bf (e)} and {\bf (f):} Increase in average EC relative to basal average EC in ascending order. {\bf (g):} Probability distribution for times of entrance into PE for the maximum (green) and minimum (red) Fano factor. {\bf (h):} Cumulative distributions for the curves in (g).}
\end{figure}
\begin{figure}
	\centering
	\includegraphics[trim=0 0 0 1.0cm, height=0.8\textheight]{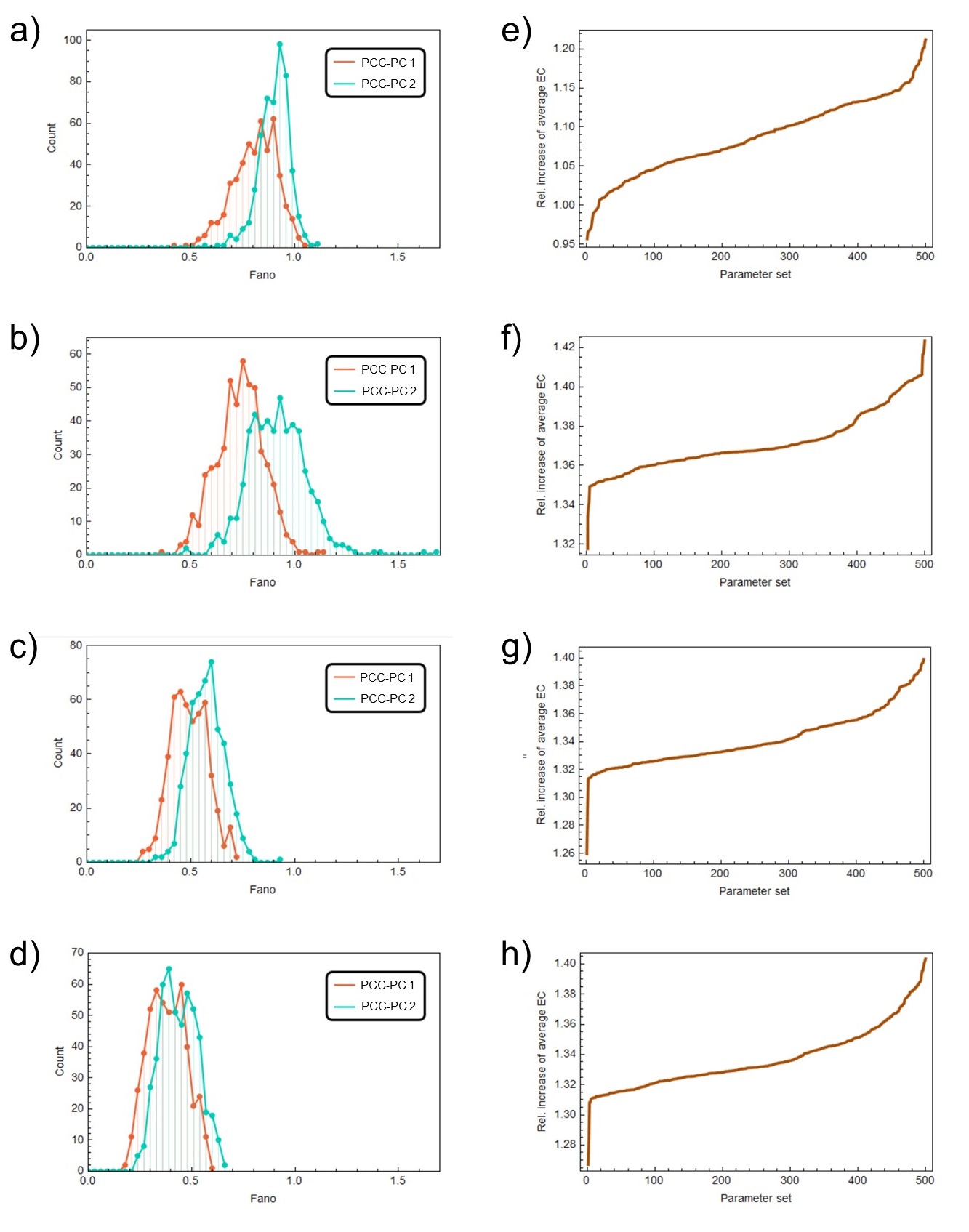}
	\caption{Comparison of results between PCC-PC interaction models 1 and 2. {\bf (a)} - {\bf (d):} Fano distributions for PCC-PC interaction models 1 (orange) and model 2 (blue). {\bf (a)} - {\bf (d):} Increase in average EC of PCC-PC interaction model 1 relative to PCC-PC interaction model 2 in ascending order.}
\end{figure}
\begin{figure}
	\centering
	\includegraphics[trim=0 0 0 1.0cm, height=0.18\textheight]{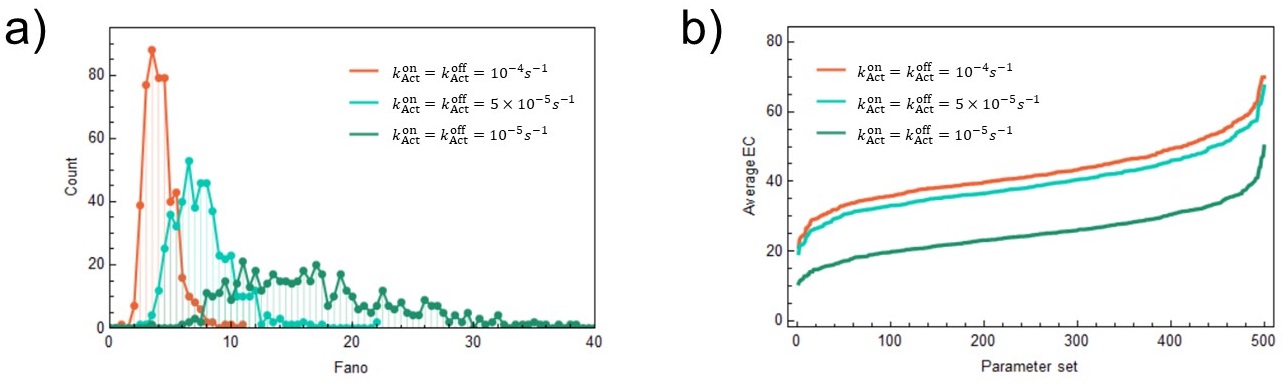}
	\caption{Fano distributions {\bf (a)} and averages of EC {\bf (b)} for $k^{\text{on}}_{\text{Act}}=k^{\text{off}}_{\text{Act}}=0$, $k^{\text{on}}_{\text{Act}}=k^{\text{off}}_{\text{Act}}=5\times 10^{-5}$s$^{-1}$ and $k^{\text{on}}_{\text{Act}}=k^{\text{off}}_{\text{Act}}=10^{-5}$s$^{-1}$.}
\end{figure}

For each of the 500 parameter sets, using the PCC-PC interaction model 1) (see previous section) a simulation was done with the following combinations of $(k^{\text{on}}_{\text{act}}, k^{\text{off}}_{\text{act}})$: $(0,0)$, $(10^{-3},10^{-3})$, 
$(10^{-3},10^{-4})$ and $(10^{-3},10^{-5})$ (in inverse seconds). The averages and Fano factors for the variable EC are presented in Fig. 5 a-d. The averages, represented by the orange dots, are sorted in an ascending order. The histograms were constructed in Mathematica using the ``Histogram" function with 30 bars. Fig. 5 e and f show the increase in average EC upon introducing an activator for different combinations $(k^{\text{on}}_{\text{act}}, k^{\text{off}}_{\text{act}})$, represented by different colors. For each color, the curves were sorted in their own ascending order. Shown in Fig. 5 g are two probability density functions $P(t)$ for the reaction Pol$^{++}_{7,5,2}$/H/Spt5$_p$/HCE/mTASE/PAFC1/SEC/BRD$\rightarrow$ EC to occur between the times $t$ and $t+dt$. Fig. 5 h shows the cumulative distribution functions $P_c(t)$ defined as
\begin{equation}\label{Cumulative}
P_c(t)=1-\int_0^tdt'P(t').
\end{equation}
The cumulative function, Eq. (\ref{Cumulative}), is used to sample $t$ by solving the equation $P_c(t)=\xi$, where $\xi$ is a random real number ranging from 0 to 1. 
The red (green) curve in both graphs was generated from a parameter set responsible for the smallest (largest) Fano factor in all the obtained data. Fig. 5 g reveals an interesting feature of this system: even though the bulk of the red curve seems to be wider than that of the green curve, the green curve has a much longer tail. If one was to sample the time at which a new EC occurred using the cumulative functions in Fig. 5 h, the times that are greater than $\sim$800 seconds would occur much more often for the green curve than the red curve.   

This entire procedure was repeated for the second PCC-PC interaction (see previous section), the results of which are shown in Fig. 6. For all combinations of $(k^{\text{on}}_{\text{act}}, k^{\text{off}}_{\text{act}})$ mentioned above, the Fano distributions tend slightly towards the right (see Fig. 6 a-d). With this increased noise, the efficiencies also increase. 
It is clear from these graphs that the distributions of EC tend to be sub-Poissonian regardless of the reaction rates. This is more true the smaller $k^{\text{off}}_{\text{act}}$ is compared to $k^{\text{on}}_{\text{act}}=10^{-3}$s$^{-1}$. 

However, what would happen if both $k^{\text{on}}_{\text{act}}$ and $k^{\text{off}}_{\text{act}}$ were small, e. g. $\sim 10^{-5}$? To answer this question, I
ran additional simulations for these combinations of $(k^{\text{on}}_{\text{act}}, k^{\text{off}}_{\text{act}})$: $(10^{-4},10^{-4})$, $(5\times10^{-5},5\times10^{-5})$ and $(10^{-5},10^{-5})$ (in inverse seconds). The results shown in Fig. 7 a) are consistent with previous research: the slower the activator dynamics, the greater the noise tends to be. The averages, shown in Fig. 7 b), do not seem to change very much compared to those in Figs. 5 and 6.

\subsection{Alternative pathways}

\begin{figure}
	\centering
	\includegraphics[trim=0 0 0 1.0cm, height=0.4\textheight]{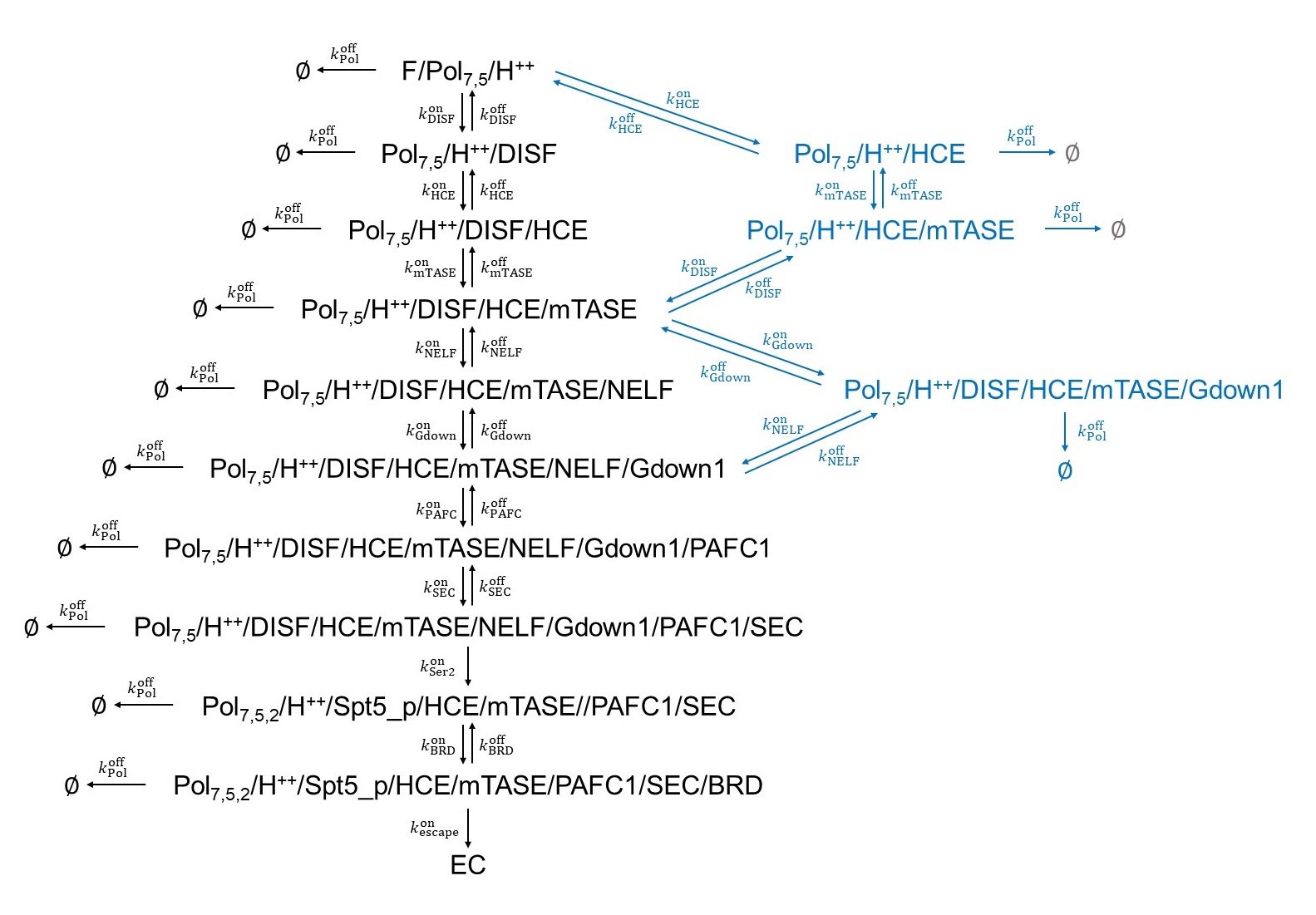}
	\caption{Two additional pathways to get from F/Pol/$_{7,5}$/H$^{++}$ to EC.}
\end{figure}
\begin{figure}
	\centering
	\includegraphics[trim=0 0 0 1.0cm, height=0.4\textheight]{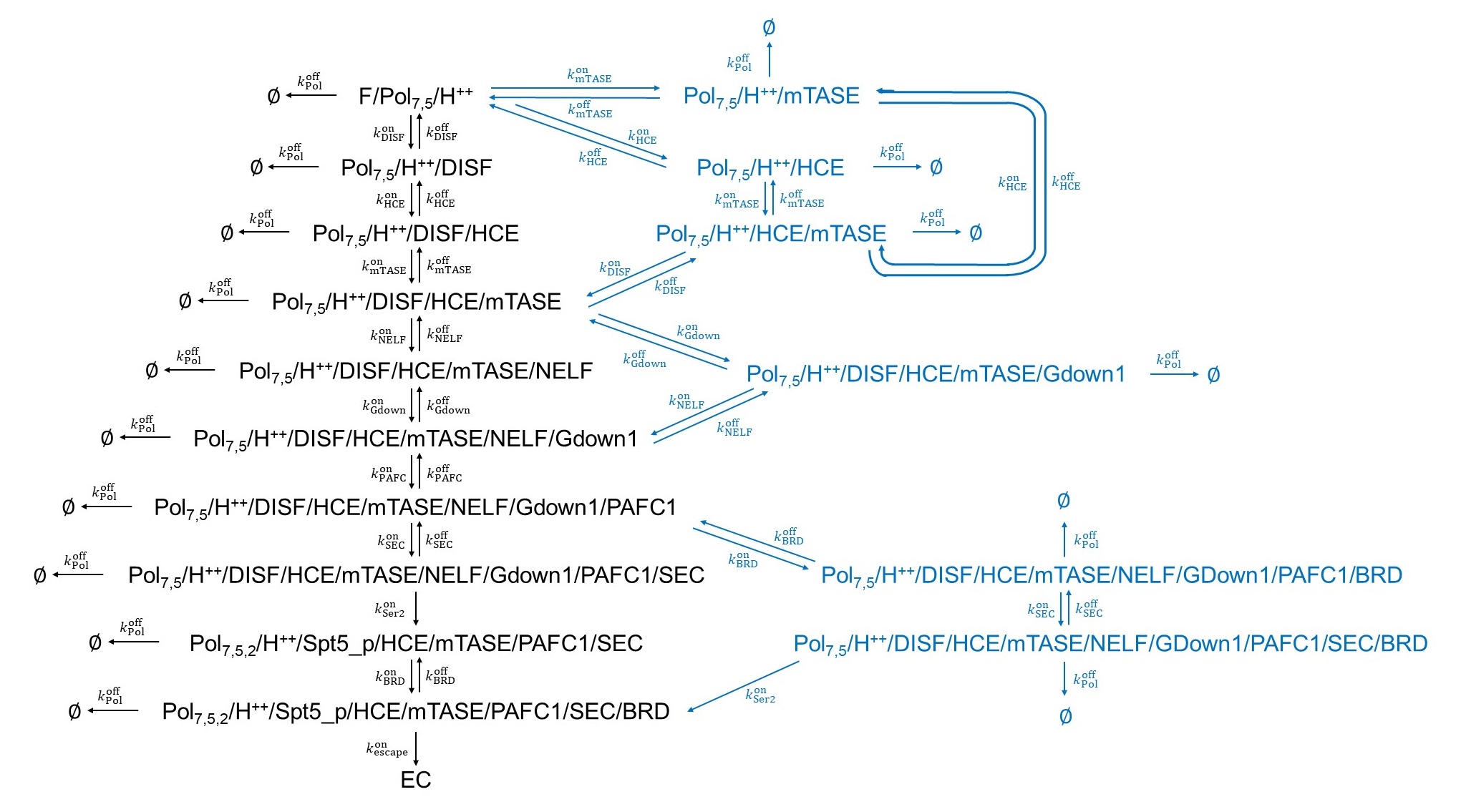}
	\caption{Four additional pathways to get from F/Pol/$_{7,5}$/H$^{++}$ to EC.}
\end{figure}

The mechanism of the PCC formation and the PPP is still a subject of ongoing research with many details to be worked out, as well as some major discoveries to be made. For example, the
precise order in which the two complexes are formed and whether they are always formed the same way has not yet reached a consensus. Given such a state of affairs, it is reasonable to
wonder if the results obtained thus far would hold up if some of the reactions in Figs. 3 and 4 could also happen in a different order. In other words, what would happen if there were alternative pathways
to get from an empty promoter to PE? To answer this question, we ran simulations for the PCC-PC interaction model 1) but with additional pathways, shown in Figs. 8 and 9.  
The reactions in blue are the additional pathways. In Fig. 8, there are two additional pathways: HCE and mTASE bind before DISF; and Gdown1 binds before NELF.
In Fig. 9, there are two additional pathways on top of those from Fig. 8: mTASE binds before HCE, followed by DISF; and BRD binds before SEC, followed by the phosphorylation of Serine 2 and Spt5.

In terms of noise, adding more pathways does not seem to change much: in Fig. 8b, the Fano distribution is slightly shifted to the right; on the other hand, in Fig. 8d, it is shifted slightly to the left. The efficiencies, however, have more of a consistent trend: adding more pathways does increase the efficiency in the ballpark of 10 to 30 percent.

\begin{figure}
	\centering
	\includegraphics[trim=0 0 0 1.0cm, height=0.8\textheight]{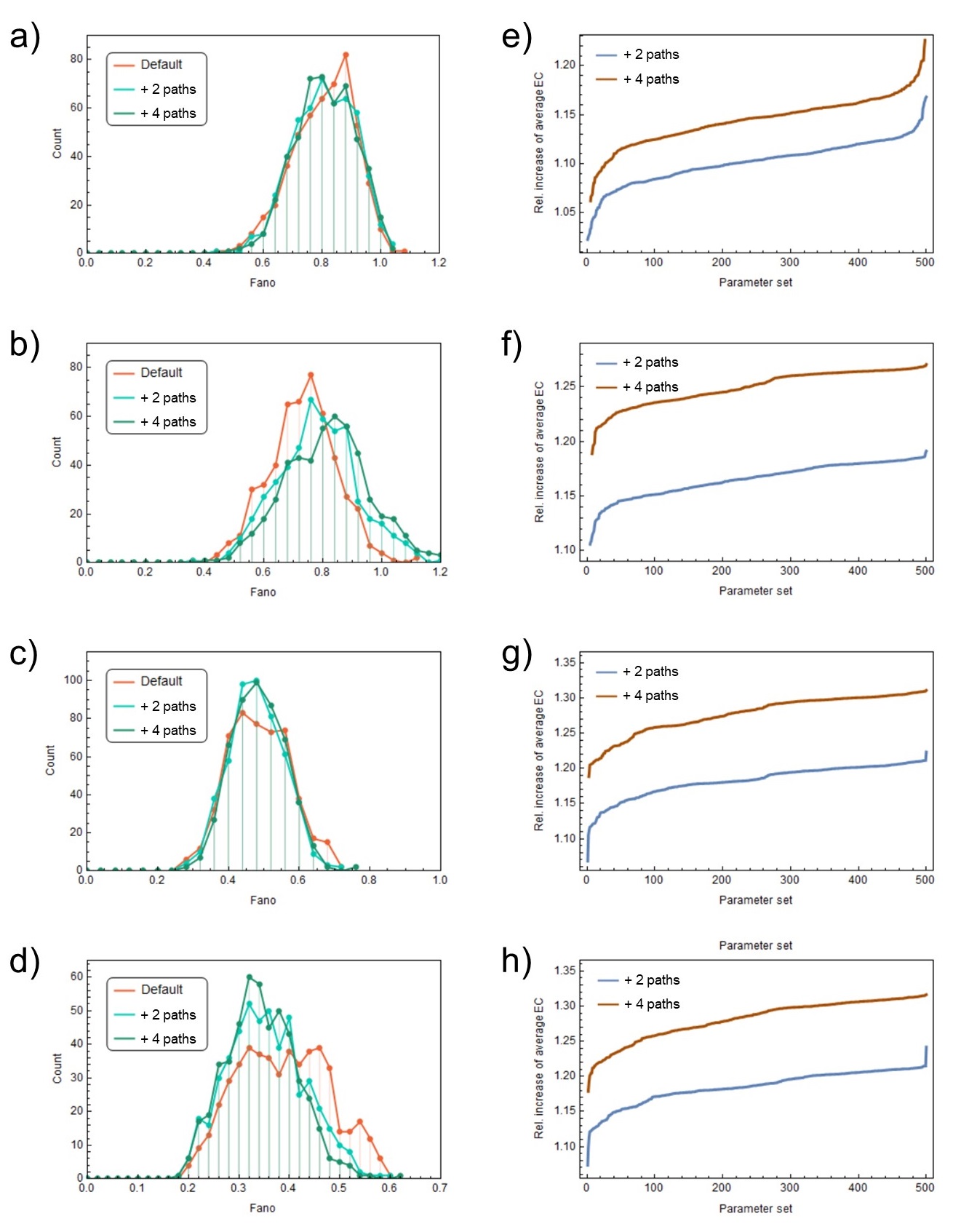}
	\caption{{\bf (a)} - {\bf (d):} Fano distributions for PCC-PC interaction model 1 (orange), two additional pathways (light blue) and four additional pathways (green). {\bf (e)} - {\bf (h):} Increase in average EC of PCC-PC interaction model 1 relative to two additional pathways (blue) and four additional pathways (brown).}
\end{figure}

%

\section*{Discussion}
Gene transcription is a process that contains several stages and thousands of reactions. Models that reduce such a complex system to a few parameters run the risk of offering erroneous
interpretations of data. In this paper we have studied stochastic dynamics of the first two stages -- the pre-initiation complex (PIC) formation and promoter proximal pausing (PPP) -- using a detailed stochastic model. The reaction rates were selected on the basis of experimental observations, either direct or inferred from optimized model parameters, and by imposing constraints on the transcriptional efficiency and on the half-life of the paused state. Four different reaction network topologies were simulated: 1) binding of Pol II was allowed only when the promoter proximal pause region was empty; 2)
Pol II was allowed to bind when the promoter proximal pause region was occupied by a Pol II complex, but it was not allowed to clear the promoter; 3) same as 1) but with two new pathways added to the reaction network; and 4) same as 1) but with four new pathways added to the reaction network.

The results showed that in most cases, regardless of the specific values of the system parameters obtained under these constraints, the number of Pol II complexes that enter productive elongation per time follow a sub-Poissonian distribution. Only 4.4 $\%$ of all cases led to super-Poissonian distributions with the maximum Fano factor of 1.7. This trend was shared by all four network topologies, although the distributions of Fano factors were shifted slightly to the right in 2). These results run contrary to the view that noise in gene expression necessarily comes from the promoter. It is certainly true that sometimes it does, as was also demonstrated herein: slow activator-enhancer dynamics led to greater variation in the number of elongation complexes. However, for fast activator-enhancer binding rates, the Fano factors remained well below 1. It is not entirely surprising that a sequential set of reactions leads to small noise; as in reaction cascades \cite{Amir} or a compartmentalization \cite{Albert0}.

This study was a step towards understanding how complexity at the promoter and promoter proximal region effectuates noise in the number of elongation complexes. Although, based on these results, one cannot comment on the fluctuations in the copy number of mRNA transcripts, as the journey to being a fully processed mRNA is far from over, it can be said that the contribution to the fluctuations coming from the PIC formation and PPP is negligible. Of course, the present study has only scratched the surface. There remain many possibilities to explore, such as: what happens when the average half-life of the PC is much longer/shorter; and how a dependence of the Pol II dissociation rates on the state of the PCC and the PC affects noise. These and other questions still need to be investigated, and this paper can serve as a bases for answering them.
\vspace{49.4mm}
\section*{Appendix}
Before we compute $\langle t_\text{clear}\rangle$, $\langle t_\text{escape}\rangle$ and $\langle t_\text{pause}\rangle$,
let us first label all the states of the PCC and PC:
\begin{eqnarray}
&&1:\,\,\,\,\,\text{TATA}\,\,\,\,\,\,\,\,\,\,\,\,\,\,\,\,\,\,\,\,\,\,\,\,\,\,\,\,\,\,\,\,\,\,\,\,\,\,\,\,\,\,\,\,\,\,\,\,\,\,\,\,\,\,\,\,\,\,\,\,\,\,\,\,\,\,\,\,\,\,\,\,9:\,\,\,\,\,\text{TATA/TBP/B/Pol$^+_7$/H}\nonumber\\
&&2:\,\,\,\,\,\text{TATA/TBP}\,\,\,\,\,\,\,\,\,\,\,\,\,\,\,\,\,\,\,\,\,\,\,\,\,\,\,\,\,\,\,\,\,\,\,\,\,\,\,\,\,\,\,\,\,\,\,\,\,\,\,\,\,\,\,\,10:\,\,\,\,\,\text{TATA/TBP/A/B/Pol$^+_{7,5}$/H}\nonumber\\
&&3:\,\,\,\,\,\text{TATA/TBP/A}\,\,\,\,\,\,\,\,\,\,\,\,\,\,\,\,\,\,\,\,\,\,\,\,\,\,\,\,\,\,\,\,\,\,\,\,\,\,\,\,\,\,\,\,\,\,\,\,\,11:\,\,\,\,\,\text{TATA/TBP/A/B/Pol}\nonumber\\
&&4:\,\,\,\,\,\text{TATA/TBP/B}\,\,\,\,\,\,\,\,\,\,\,\,\,\,\,\,\,\,\,\,\,\,\,\,\,\,\,\,\,\,\,\,\,\,\,\,\,\,\,\,\,\,\,\,\,\,\,\,\,12:\,\,\,\,\,\text{TATA/TBP/A/B/Pol/H}\nonumber\\
&&5:\,\,\,\,\,\text{TATA/TBP/A/B}\,\,\,\,\,\,\,\,\,\,\,\,\,\,\,\,\,\,\,\,\,\,\,\,\,\,\,\,\,\,\,\,\,\,\,\,\,\,\,\,\,\,13:\,\,\,\,\,\text{TATA/TBP/A/B/Pol$^+$/H}\nonumber\\
&&6:\,\,\,\,\,\text{TATA/TBP/B/Pol}\,\,\,\,\,\,\,\,\,\,\,\,\,\,\,\,\,\,\,\,\,\,\,\,\,\,\,\,\,\,\,\,\,\,\,\,\,\,14:\,\,\,\,\,\text{TATA/TBP/A/B/Pol$^+_7$/H}\nonumber\\
&&7:\,\,\,\,\,\text{TATA/TBP/B/Pol/H}\,\,\,\,\,\,\,\,\,\,\,\,\,\,\,\,\,\,\,\,\,\,\,\,\,\,\,\,\,\,\,15:\,\,\,\,\,\text{TATA/TBP/A/B/Pol$^+_{7,5}$/H}\nonumber\\
&&8:\,\,\,\,\,\text{TATA/TBP/B/Pol$^+$/H}\nonumber
\end{eqnarray}
and
\begin{eqnarray}
&&1:\,\,\,\,\,\text{F/Pol$_{7,5}$/H$^{++}$}\nonumber\\
&&2:\,\,\,\,\,\text{F/Pol$_{7,5}$/H$^{++}$/DISF}\nonumber\\
&&3:\,\,\,\,\,\text{F/Pol$_{7,5}$/H$^{++}$/DISF/HCE}\nonumber\\
&&4:\,\,\,\,\,\text{F/Pol$_{7,5}$/H$^{++}$/DISF/HCE/mTASE}\nonumber\\
&&5:\,\,\,\,\,\text{F/Pol$_{7,5}$/H$^{++}$/DISF/HCE/mTASE/NELF}\nonumber\\
&&6:\,\,\,\,\,\text{F/Pol$_{7,5}$/H$^{++}$/DISF/HCE/mTASE/NELF/Gdown1}\nonumber\\
&&7:\,\,\,\,\,\text{F/Pol$_{7,5}$/H$^{++}$/DISF/HCE/mTASE/NELF/Gdown1/PAFC1}\nonumber\\
&&8:\,\,\,\,\,\text{F/Pol$_{7,5}$/H$^{++}$/DISF/HCE/mTASE/NELF/Gdown1/PAFC1/SEC}\nonumber\\
&&9:\,\,\,\,\,\text{F/Pol$_{7,5,2}$/H$^{++}$/Spt5$_p$/HCE/mTASE/NELF/Gdown1/PAFC1/SEC}\nonumber\\
&&10:\,\,\,\,\,\text{F/Pol$_{7,5,2}$/H$^{++}$/Spt5$_p$/HCE/mTASE/NELF/Gdown1/PAFC1/SEC/BRD}\nonumber
\end{eqnarray}
To compute $\langle t_\text{clear}\rangle$, we need to know the cumulative probability $Q(t)$ that either of these reactions
\begin{eqnarray}
	&&\text{R$_1$:\,\,\,\,\,\,TATA/TBP/B/Pol$_{7,5}^{+}$/H$\rightarrow$F/Pol$_{7,5}$/H$^{++}$+TATA/TBP}\nonumber\\
	&&\text{R$_2$:\,\,\,\,\,\,TATA/TBP/A/B/Pol$_{7,5}^{+}$/H$\rightarrow$F/Pol$_{7,5}$/H$^{++}$+TATA/TBP/A}\nonumber\\
\end{eqnarray}
will occur at time $t$. The average of times sampled from $Q(t)$, i. e. $\langle t_\text{clear}\rangle$, is equal to
\begin{equation}
\langle t_\text{clear}\rangle=\int_0^{\infty}dt'Q(t').
\end{equation}
To compute $Q(t)$, let us first define a system $S'$ that is identical to the system in Fig. 3 but one which does not
contain reactions R$_1$ and R$_2$. Then, $Q(t)$ can be computed by a previously published method \cite{Albert,Albert2}, which states that
\begin{equation}
	Q(t)=\sum_iQ_i(t),
\end{equation}
where $Q_i(t)$ satisfy
\begin{equation}\label{Qi}
	\frac{d}{dt}Q_i(t)=\sum_j[T_{ij}+M_{ij}]Q_j(t).
\end{equation}
The nondiagonal matrix elements of $T_{ij}$
\small
\begin{eqnarray}
	T_{i\neq j}=\left[\begin{array}{ccccccccccccccc}
		\cdot & k^{\text{off}}_{\text{TBP}} & 0 & 0 & 0 & 0 & 0 & 0 & 0 & 0 & 0 & 0 & 0 & 0 & 0 \\
		
		k^{\text{on}}_{\text{TBP}} & \cdot & k^{\text{off}}_{\text{A}} & k^{\text{off}}_{\text{B}} & 0 & 0 & 0 & k^{\text{off}}_{\text{Pol}} & k^{\text{off}}_{\text{Pol}} & k^{\text{off}}_{\text{Pol}} & 0 & 0 & 0 & 0 & 0  \\
		
		0 & k^{\text{on}}_{\text{A}} & \cdot & 0 & k^{\text{off}}_{\text{A$|$B}} & 0 & 0 & 0 & 0 & 0 & 0 & 0 & k^{\text{off}}_{\text{Pol}} & k^{\text{off}}_{\text{Pol}} & k^{\text{off}}_{\text{Pol}}\\
		
		0 & k^{\text{on}}_{\text{B}} & 0 & \cdot & k^{\text{off}}_{\text{A}} & k^{\text{off}}_{\text{Pol}} & k^{\text{off}}_{\text{Pol}} & 0 & 0 & 0 & 0 & 0 & 0 & 0 & 0  \\
		
		0 & 0 & k^{\text{on}}_{\text{B}} & k^{\text{on}}_{\text{A}} & \cdot & 0 & 0 & 0 & 0 & 0 & k^{\text{off}}_{\text{Pol}} & k^{\text{off}}_{\text{Pol}} & 0 & 0 & 0  \\
		
		0 & 0 & 0 & k^{\text{on}}_{\text{Pol}} & 0 & \cdot &  k^{\text{off}}_{\text{H}} & 0 & 0 & 0 &  k^{\text{off}}_{\text{A}} & 0 & 0 & 0 & 0  \\
		
		0 & 0 & 0 & 0 & 0 & k^{\text{on}}_{\text{H}} & \cdot & 0 & 0 & 0 & 0 & k^{\text{off}}_{\text{A}} & 0 & 0 & 0  \\
		
		0 & 0 & 0 & 0 & 0 & 0 & k^{\text{on}}_{\text{+}} & \cdot & 0 & 0 & 0 & 0 & k^{\text{off}}_{\text{A}} & 0 & 0  \\
		
		0 & 0 & 0 & 0 & 0 & 0 & 0 & k^{\text{on}}_{\text{Ser7}} & \cdot & 0 & 0 & 0 & 0 & k^{\text{off}}_{\text{A}} & 0  \\
		
		0 & 0 & 0 & 0 & 0 & 0 & 0 & 0 & k^{\text{on}}_{\text{Ser5}} & \cdot & 0 & 0 & 0 & 0 & k^{\text{off}}_{\text{A}}  \\
		
		0 & 0 & 0 & 0 & 0 & 0 & k^{\text{on}}_{\text{A}} & 0 & 0 & 0 & k^{\text{on}}_{\text{H}} & \cdot & 0 & 0 & 0  \\
		
		0 & 0 & 0 & 0 & 0 & 0 & 0 & k^{\text{on}}_{\text{A}} & 0 & 0 & 0 & k^{\text{on}}_{\text{+}} & \cdot & 0 & 0  \\
		
		0 & 0 & 0 & 0 & 0 & 0 & 0 & 0 & k^{\text{on}}_{\text{A}} & 0 & 0 & 0 & k^{\text{on}}_{\text{Ser7}} & \cdot & 0  \\
		
		0 & 0 & 0 & 0 & 0 & 0 & 0 & 0 & 0 & k^{\text{on}}_{\text{A}} & 0 & 0 & 0 & k^{\text{on}}_{\text{Ser5}} & \cdot  \\
		
	\end{array}\right]\nonumber
\end{eqnarray}
\normalsize
are the probabilities for state $j\neq i$ to transition to state $i$; the diagonal elements, $T_{ii}$ 
\begin{eqnarray}
	&&T_{11}=-k^{\text{on}}_{\text{TBP}}\,\,\,\,\,\,\,\,\,\,\,\,\,\,\,\,\,\,\,\,\,\,\,\,\,\,\,\,\,\,\,\,\,\,\,\,\,\,\,\,\,\,\,\,\,\,\,\,\,\,\,\,\,\,\,\,\,\,\,\,\,\,\,\,\,\,\,\,\,\,T_{99}=-(k^{\text{on}}_{\text{A}}+k^{\text{on}}_{\text{Ser5}}+k^{\text{off}}_{\text{Pol}})\nonumber\\
	&&T_{22}=-(k^{\text{on}}_{\text{A}}+k^{\text{on}}_{\text{B}}+k^{\text{off}}_{\text{TBP}})\,\,\,\,\,\,\,\,\,\,\,\,\,\,\,\,\,\,\,\,\,\,\,\,\,\,\,\,\,\,\,\,\,\,\,T_{10,10}=-(k^{\text{on}}_{\text{A}}+k^{\text{off}}_{\text{Pol}})\nonumber\\
	&&T_{33}=-(k^{\text{off}}_{\text{A}}+k^{\text{on}}_{\text{A$|$B}})\,\,\,\,\,\,\,\,\,\,\,\,\,\,\,\,\,\,\,\,\,\,\,\,\,\,\,\,\,\,\,\,\,\,\,\,\,\,\,\,\,\,\,\,\,\,\,\,\,\,\,\,T_{11,11}=-(k^{\text{on}}_{\text{A}}+k^{\text{on}}_{\text{H}}+k^{\text{off}}_{\text{Pol}})\nonumber\\
	&&T_{44}=-(k^{\text{on}}_{\text{A}}+k^{\text{off}}_{\text{B}}+k^{\text{on}}_{\text{Pol}})\,\,\,\,\,\,\,\,\,\,\,\,\,\,\,\,\,\,\,\,\,\,\,\,\,\,\,\,\,\,\,\,\,\,\,\,\,\,T_{12,12}=-(k^{\text{off}}_{\text{A}}+k^{\text{off}}_{\text{H}}+k^{\text{on}}_{\text{+}}+k^{\text{off}}_{\text{Pol}})\nonumber\\
	&&T_{55}=-(k^{\text{off}}_{\text{A}}+k^{\text{off}}_{\text{A$|$B}}+k^{\text{on}}_{\text{Pol}})\,\,\,\,\,\,\,\,\,\,\,\,\,\,\,\,\,\,\,\,\,\,\,\,\,\,\,\,\,\,\,\,\,\,\,T_{13,13}=-(k^{\text{off}}_{\text{A}}+k^{\text{on}}_{\text{Ser7}}+k^{\text{off}}_{\text{Pol}})\nonumber\\
	&&T_{66}=-(k^{\text{on}}_{\text{A}}+k^{\text{on}}_{\text{H}}+k^{\text{off}}_{\text{Pol}})\,\,\,\,\,\,\,\,\,\,\,\,\,\,\,\,\,\,\,\,\,\,\,\,\,\,\,\,\,\,\,\,\,\,\,\,\,\,T_{14,14}=-(k^{\text{off}}_{\text{A}}+k^{\text{on}}_{\text{Ser5}}+k^{\text{off}}_{\text{Pol}})\nonumber\\
	&&T_{77}=-(k^{\text{on}}_{\text{A}}+k^{\text{on}}_{\text{H}}+k^{\text{on}}_{\text{+}}+k^{\text{off}}_{\text{Pol}})\,\,\,\,\,\,\,\,\,\,\,\,\,\,\,\,\,\,\,\,\,\,\,T_{15,15}=-(k^{\text{off}}_{\text{A}}+k^{\text{off}}_{\text{Pol}}),\nonumber\\
	&&T_{88}=-(k^{\text{on}}_{\text{A}}+k^{\text{on}}_{\text{Ser7}}+k^{\text{off}}_{\text{Pol}})\nonumber\\
\end{eqnarray}
are probabilities for state $i$ to
transition out of $i$. The always diagonal matrix elements $M_{ij}=k^{\text{on}}_{\text{clear}}\delta_{ij}(\delta_{i,10}+\delta_{i,15})$ give the probability for a state $i$ to transition into F/Pol$_{7,5}$/H$^{++}$ via R$_1$ or R$_2$.
If all elements of $M_{ij}$ were zero, Eq. (\ref{Qi}) would become the Master equation for $S'$.

The same method can be applied to compute $\langle t_\text{escape}\rangle$. In this case, $S'$ contains all reactions in Fig. 4 except for the one that leads to PE. When a reaction that leads to the state
$\emptyset$ occurs, i. e. where the Pol II complex dissociates from the DNA, the variable F/Pol$_{7,5}$/H$^{++}$ is set to 1, and the process of reaching PE must start all over again. 
The off-diagonal elements of $T_{ij}$ for this system then read:
\small
\begin{eqnarray}\label{ToffD}
	T_{i\neq j}=\left[\begin{array}{cccccccccc}
		\cdot & k^{\text{off}}_{\text{DISF}} & k^{\text{off}}_{\text{Pol}} & k^{\text{off}}_{\text{Pol}} & k^{\text{off}}_{\text{Pol}} & k^{\text{off}}_{\text{Pol}} & k^{\text{off}}_{\text{Pol}} & k^{\text{off}}_{\text{Pol}} & k^{\text{off}}_{\text{Pol}} & k^{\text{off}}_{\text{Pol}} \\
		
		k^{\text{on}}_{\text{DISF}} & \cdot & k^{\text{off}}_{\text{HCE}} & 0 & 0 & 0 & 0 & 0 & 0 & 0 \\
	
		0 & k^{\text{on}}_{\text{HCE}} & \cdot & k^{\text{off}}_{\text{mTASE}} & 0 & 0 & 0 & 0 & 0 & 0 \\
		
		0 & 0 & k^{\text{on}}_{\text{mTASE}} & \cdot & k^{\text{off}}_{\text{NELF}} & 0 & 0 & 0 & 0 & 0 \\
		
		0 & 0 & 0 & k^{\text{on}}_{\text{NELF}} & \cdot & k^{\text{off}}_{\text{Gdown}} & 0 & 0 & 0 & 0 \\
		
		0 & 0 & 0 & 0 & k^{\text{on}}_{\text{Gdown}} & \cdot & k^{\text{off}}_{\text{PAFC}} & 0 & 0 & 0 \\
		
		0 & 0 & 0 & 0 & 0 & k^{\text{on}}_{\text{PAFC}} & \cdot & k^{\text{off}}_{\text{SEC}} & 0 & 0 \\
		
		0 & 0 & 0 & 0 & 0 & 0 & k^{\text{on}}_{\text{SEC}} & \cdot & 0 & 0 \\
		
		0 & 0 & 0 & 0 & 0 & 0 & 0 & k^{\text{on}}_{\text{Ser2}} & \cdot & k^{\text{off}}_{\text{BRD}} \\
		
		0 & 0 & 0 & 0 & 0 & 0 & 0 & 0 & k^{\text{on}}_{\text{BRD}} & \cdot \\
				
	\end{array}\right],
\end{eqnarray}
\normalsize
while the diagonal elements are given by
\begin{eqnarray}\label{TD}
	&&T_{11}=-k^{\text{on}}_{\text{DISF}}\,\,\,\,\,\,\,\,\,\,\,\,\,\,\,\,\,\,\,\,\,\,\,\,\,\,\,\,\,\,\,\,\,\,\,\,\,\,\,\,\,\,\,\,\,\,\,\,\,\,\,\,\,\,\,\,\,\,\,\,\,\,\,\,\,\,\,\,T_{66}=-(k^{\text{off}}_{\text{Gdown}}+k^{\text{on}}_{\text{PAFC}}+k^{\text{off}}_{\text{Pol}})\nonumber\\
	&&T_{22}=-(k^{\text{off}}_{\text{DISF}}+k^{\text{on}}_{\text{HCE}}+k^{\text{off}}_{\text{Pol}})\,\,\,\,\,\,\,\,\,\,\,\,\,\,\,\,\,\,\,\,\,\,\,\,\,\,\,T_{77}=-(k^{\text{off}}_{\text{PAFC}}+k^{\text{on}}_{\text{SEC}}+k^{\text{off}}_{\text{Pol}})\nonumber\\
	&&T_{33}=-(k^{\text{off}}_{\text{HCE}}+k^{\text{on}}_{\text{mTASE}}+k^{\text{off}}_{\text{Pol}})\,\,\,\,\,\,\,\,\,\,\,\,\,\,\,\,\,\,\,\,\,\,T_{88}=-(k^{\text{off}}_{\text{SEC}}+k^{\text{on}}_{\text{Ser2}}+k^{\text{off}}_{\text{Pol}})\nonumber\\
	&&T_{44}=-(k^{\text{off}}_{\text{mTASE}}+k^{\text{on}}_{\text{NELF}}+k^{\text{off}}_{\text{Pol}})\,\,\,\,\,\,\,\,\,\,\,\,\,\,\,\,\,\,\,\,T_{99}=-(k^{\text{on}}_{\text{BRD}}+k^{\text{off}}_{\text{Pol}})\nonumber\\
	&&T_{55}=-(k^{\text{off}}_{\text{NELF}}+k^{\text{off}}_{\text{Gdown}}+k^{\text{off}}_{\text{Pol}})\,\,\,\,\,\,\,\,\,\,\,\,\,\,\,\,\,\,\,\,\,T_{10,10}=-(k^{\text{off}}_{\text{BRD}}+k^{\text{off}}_{\text{Pol}}).\nonumber
\end{eqnarray}

For the case of $\langle t_\text{pause}\rangle$, we need to remove all reactions that lead to the state $\emptyset$, and also the reaction that leads to PE.
This renders 
both diagonal and off-diagonal elements of $T_{ij}$ identical to Eq. (\ref{ToffD}) and (\ref{TD}) at $k^{\text{off}}_{\text{Pol}}=0$, while
$M_{ij}=-[k^{\text{off}}_{\text{Pol}}(1-\delta_{1,i})+k_{\text{escape}}^{\text{on}}\delta_{i,10}]\delta_{ij}$. The reason for omitting the state $i=1$ from $M_{ij}$ is that via the
reaction F/Pol$_{7,5}$/H$^{++}\rightarrow \emptyset$ it leads to itself (since $\emptyset=$F/Pol$_{7,5}$/H$^{++}$).


\section*{Acknowledgments}

We want to thank the creator of the YouTube channel theCrux \\
(https://www.youtube.com/c/theCrux/featured).
It was a starting point for researching material for this paper.

\end{document}